\documentclass[aps,pra,preprint,showpacs]{revtex4}
\usepackage{graphicx}
\def\C{{\bf C}}
\def\Z{{\bf Z}}

\begin{document}

\title{Large Qudit Limit of One-dimensional Quantum Walks}

\author{Mitsunori Sato}
%\email[]{}
\affiliation{Department of Physics,
Faculty of Science and Engineering,
Chuo University, Kasuga, Bunkyo-ku, Tokyo 112-8551, Japan}
\author{Naoki Kobayashi}
\email[]{knaoki@phys.chuo-u.ac.jp}
\affiliation{Department of Physics,
Faculty of Science and Engineering,
Chuo University, Kasuga, Bunkyo-ku, Tokyo 112-8551, Japan}
\author{Makoto Katori} 
\email[]{katori@phys.chuo-u.ac.jp}
\affiliation{Department of Physics,
Faculty of Science and Engineering,
Chuo University, Kasuga, Bunkyo-ku, Tokyo 112-8551, Japan}
\author{Norio Konno}
\email[]{konno@ynu.ac.jp}
\affiliation{
Department of Applied Mathematics, 
Yokohama National University, 
79-5 Tokiwadai, Yokohama 240-8501, Japan}

%\date{\today}
\date{14 February 2008}

\begin{abstract}
We study a series of one-dimensional 
discrete-time quantum-walk models 
labeled by half integers
$j=1/2, 1, 3/2, \cdots$, introduced by
Miyazaki {\it et al.}, each of
which the walker's wave function has $2j+1$ 
components and hopping range at each 
time step is $2j$.
In long-time limit the density functions of 
pseudovelocity-distributions are generally given 
by superposition of
appropriately scaled Konno's density function.
Since Konno's density function has a finite open
support and it diverges at the boundaries of 
support, limit distribution of pseudovelocities
in the $(2j+1)$-component model can have $2j+1$ pikes,
when $2j+1$ is even. 
When $j$ becomes very large, however, we found that 
these pikes vanish and a universal and monotone
convex structure appears around the origin in limit distributions.
We discuss a possible route from 
quantum walks to classical diffusion 
associated with the $j \to \infty$ limit.
\end{abstract}

% insert suggested PACS numbers in braces on next line
% insert suggested keywords - APS authors don't need to do this
%\keywords{}

\maketitle

%%%%%%%%%%%%%%%%%%%%%%%%%%%%%%%%%%%%%%%%%%%%%%%%%%%%%%%%%%%%%%%%%%%%%%
\section{introduction}
%%%%%%%%%%%%%%%%%%%%%%%%%%%%%%%%%%%%%%%%%%%%%%%%%%%%%%%%%%%%%%%%%%%%%%

In contrast with diffusive behavior of classical
random walks, in which the standard deviation
of walker's positions from the starting point
is proportional to the square root of time-step
in the long-time limit,
quantum walkers have velocities \cite{ADZ93,Mey96,NV00,ABNVW01,
TM02,Kem03,Amb03,BCA03,Ken06}, 
and probability distributions of pseudovelocity,
the position divided by time, are discussed
\cite{Kon02,Kon05,GJS04,KFK05,MKK07,Kon07}.

In the present paper we consider a series of discrete-time
quantum-walk models on the one-dimensional lattice 
({\it i.e.} integers) $\Z=\{\cdots,-2,-1,0,1,2,\cdots\}$, 
introduced by \cite{MKK07},
where models are labeled by half integers
$j=1/2,1,3/2,\cdots$.
In the model indexed by $j$, the state of quantum walker is
described by a $(2j+1)$-component wave function
$$
\Psi^{(j)}(x,t)
=\left( \begin{array}{c}
\psi_{j}^{(j)}(x,t) \cr \psi_{j-1}^{(j)}(x,t) \cr \cdots \cr 
\psi_{-j+1}^{(j)}(x,t) \cr \psi_{-j}^{(j)}(x,t)
\end{array} \right),
\quad x\in \Z, \quad
t=0,1,2,\dots, 
$$
normalized as $\sum_{x\in \Z}^{}|\Psi^{(j)}(x,t)|^{2}=1$,
where $|\Psi^{(j)}(x,t)|^{2}=[\Psi^{(j)}(x,t)]^{\dag}\Psi^{(j)}(x,t)
=\sum_{m=-j}^{j}|\psi^{(j)}_{m}(x,t)|^{2}$.
Let $R^{(j)}$ be a quantum coin represented by a
$(2j+1)\times(2j+1)$ unitary matrix, whose $(m, m^{\prime})$-component
is denoted by $R^{(j)}_{m m^{\prime}}$.
In the present paper, when we write matrices 
and vectors whose elements are
labeled by $m, m^{\prime}$, we will 
assume that the indices $m$ and $m^{\prime}$
run from $j$ to $-j$ in step of $-1$. 
At each time step, $2j+1$ components of wave function is mixed by
the quantum-coin matrix $R^{(j)}$, and 
then the quantum walker hops to $2j+1$ sites,
\begin{eqnarray}
\psi^{(j)}_{m}(x,t+1)&=&\sum_{m^{\prime}=-j}^{j}
R^{(j)}_{mm^{\prime}}\psi^{(j)}_{m^{\prime}}(x+2m,t),
\quad t=0,1,2 \cdots.
\label{eqn:hopp} 
\end{eqnarray}
We use Wigner's rotation matrices \cite{Wig59,Mes62} 
specified by three real
parameters called the Euler angles $\alpha,\beta$, and $\gamma$ as 
the quantum-coin matrix,
$$
R^{(j)}_{m m^{\prime}}(\alpha,\beta,\gamma)=
e^{-i\alpha m}r^{(j)}_{m m^{\prime}}(\beta)
e^{-i\gamma m^{\prime}},
\quad -j \leq m, m^{\prime} \leq j
$$
with
$$
r^{(j)}_{m m^{\prime}}(\beta)
= \sum_{\ell} \Gamma(j, m, m^{\prime}, \ell)
\left(\cos \frac{\beta}{2} \right)^{2j+m-m^{\prime}-2 \ell}
\left(\sin \frac{\beta}{2} \right)^{2\ell+m^{\prime}-m}.\nonumber
$$
Here
$$
\Gamma(j, m, m', \ell)=(-1)^{\ell}
\frac{\sqrt{(j+m)! (j-m)! (j+m')! (j-m')!}}
{(j-m'-\ell)! (j+m-\ell)! \ell ! (\ell +m'-m)!},\nonumber
$$
and the summation $\sum_{\ell}^{}$ extends over all integers
$\ell$, for which the arguments of the factorials
are positive or null ($0! \equiv 1$).
The position of the quantum walker at time $t$ is 
denoted by $X^{(j)}_{t}$
and the probability to find a walker
at site $x$ at time $t$ is given by 
\begin{eqnarray}
P^{(j)}(x,t) \equiv 
{\rm Prob}(X_{t}^{(j)}=x)=|\Psi^{(j)}(x,t)|^{2}.\nonumber
\end{eqnarray}
The ratio $X^{(j)}_{t}/t$ is called 
the pseudovelocity of walker \cite{KFK05}
and its $r$-th moment is given as
$$
\left\langle \left( \frac{X^{(j)}_{t}}{t} \right)^{r} \right\rangle
= \sum_{x\in \Z}^{} \left(\frac{x}{t} \right)^{r}
P^{(j)}(x,t)\nonumber
$$
at each time $t=0,1,2,\cdots$.

For simplicity, we will assume that at the initial time $t=0$
one quantum walker exists at the origin,
$$
\Psi^{(j)}(x,0)=\phi_{0}^{(j)}\delta_{x, 0}\nonumber
$$
with
\begin{equation}
\phi_{0}^{(j)}={}^{T}(q_{j},q_{j-1},\cdots, q_{-j+1},q_{-j}), 
\label{eqn:initial}
\end{equation}
where $q_{m}\in \C \equiv$ the set of all complex numbers, 
$-j \leq m \leq j$, with
$\sum_{m=-j}^{j}|q_{m}|^{2}=1$.
In this paper 
the left-superscript $T$ denotes the transpose of vector or matrix.
We usually call $\phi_{0}^{(j)}$ a $(2j+1)$-component qudit,
which the quantum walker possesses at $t=0$,

For $|a|\leq 1$, let 
\begin{equation}
\mu(x; a)=\frac{\sqrt{1-a^2}}
{\pi (1-x^2) \sqrt{a^2-x^2}}
{\bf 1}_{\{|x| < |a|\}},
\label{eqn:Konno}
\end{equation} 
where ${\bf 1}_{\{\omega\}}$
is the indicator function of a condition $\omega$; ${\bf 1}_{\{\omega\}}=1$
if the condition $\omega$ is satisfied,
${\bf 1}_{\{\omega\}}=0$ otherwise.
We call it Konno's density function,
since it was first introduced by Konno
to describe the limit distributions of the
standard two-component quantum walks
in his weak limit-theorem \cite{Kon02,Kon05}.
In an earlier paper \cite{MKK07}, the following limit theorem 
was proved,
$$
\lim_{t \to \infty} 
\left\langle \left( \frac{X^{(j)}_{t}}{t} \right)^{r}
\right\rangle
= \int_{-\infty}^{\infty} dv \, v^{r}
\nu^{(j)}(v), \quad r=0,1,2, \cdots 
$$
with
\begin{equation}
\nu^{(j)}(v)= \sum_{m: 0 < m \leq j}
\frac{1}{2m} \mu \left( \frac{v}{2m}; \cos \frac{\beta}{2} \right)
{\cal M}^{(j,m)}\left(\frac{v}{2m}\right)
+{\bf 1}_{\{(2j+1) \, \mbox{is odd}\}}
\Delta^{(j)} \delta(v),
\label{eqn:dist}
\end{equation}
where ${\cal M}^{(j,m)}(x)$ are polynomials
of $x$ of order $2j$. 
That is, the long-time limit distribution of pseudovelocity of
quantum walker is described by superposition of appropriately
scaled Konno's density functions (\ref{eqn:Konno}) with weight functions 
${\cal M}^{(j,m)}(v/2m)$, and a point mass at the 
origin with intensity $\Delta^{(j)}$, 
if the number of states $(2j+1)$ is odd.
In the previous paper \cite{MKK07}, however, 
explicit expressions for 
the weight functions
${\cal M}^{(j,m)}(x)$ are given only for $j=1/2,1,$ and $3/2$,
since the functions seem to become very complicated 
as the value of $j$ increases.

In the present paper we will calculate ${\cal M}^{(j,m)}(x)$ for
large values of $j$ and study the asymptotics of 
the limit distributions (\ref{eqn:dist}) in the $j\rightarrow\infty$ limit.
We will report our observation that, when $j$ becomes very large,
complicated structures of the limit distributions are smeared out 
and a universal monotonic convex-structure 
appears around the origin.
This observation leads us to a discussion
on a possible route
from quantum walks to classical diffusion.
Relationship between
the quantum-walk behavior and 
diffusive behavior of classical random-walk
is an important topic in the study of
quantum walks \cite{BCA03b}.

This paper is organized as follows.
In Sec.II, hermitian matrix-representations of
weight functions ${\bf M}^{(j,m)}(x)$ are introduced
and formulas are given for the matrix elements,
which are useful to calculate the limit
density-functions for large values of $j$.
We analyze limit density-functions for
large $j$ in Sec.III and
clarify the properties of convex structure,
which appears around the origin in limit
distributions for sufficiently large values of $j$.
Crossover phenomenon from quantum walks to
classical diffusion associated with the $j \to \infty$ limit
is discussed in Sec.IV.
Appendices are used for some 
details of calculations.
%%%%%%%%%%%%%%%%%%%%%%%%%%%%%%%%%%%%%%%%%%%%%%%%%%%%%%%%%%%%%%%%%%%%%%
\section{Weight Functions}
%%%%%%%%%%%%%%%%%%%%%%%%%%%%%%%%%%%%%%%%%%%%%%%%%%%%%%%%%%%%%%%%%%%%%%
\subsection{Hermitian-matrix representations}
%%%%%%%%%%%%%%%%%%%%%%%%%%%%%%%%%%%%%%%%%%%%%%%%%%%%%%%%%%%%%%%%%%%%%%
We note that the weight functions ${\cal M}^{(j,m)}(x)$ are represented
using $(2j+1)\times(2j+1)$ hermitian matrices ${\bf M}^{(j,m)}(x)$
and $(2j+1)$-component initial-qudit (\ref{eqn:initial}) as
$$
{\cal M}^{(j,m)}(x)=
[\phi_{0}^{(j)}]^{\dag}[{\bf M}^{(j,m)}(x)]\phi_{0}^{(j)}. 
$$
In Appendix A, we give the matrices ${\bf M}^{(j,m)}(x)=
({\bf M}^{(j,m)}_{m_{1}m_{2}}(x))$ for $j=1/2,1,3/2$
as examples.
In general,
\begin{eqnarray}
\overline{\bf M}^{(j,m)}_{m_{2} m_{1}}={\bf M}^{(j,m)}_{m_{1} m_{2}},
\quad -j \leq m_1, m_2 \leq j,
\quad \mbox{(hermitian condition)}
\label{eqn:hermitian}
\end{eqnarray}
and 
\begin{eqnarray}
{\bf M}^{(j,m)}_{-m_{2} -m_{1}}(x)&=&
(-1)^{m_{1}+m_{2}+2m} {\bf M}^{(j,m)}_{m_{1} m_{2}}(-x),
\quad -j \leq m_1, m_2 \leq j.
\label{eqn:symmetry}
\end{eqnarray}
We found that, if the indices $m_{1}$ and $m_{2}$
satisfy the condition 
\begin{equation}
m_1 \leq m_2 \quad
\mbox{and} \quad
m_1 \geq -m_2,
\label{eqn:condition1}
\end{equation}
we have the expression
\begin{eqnarray}
&& {\bf M}^{(j,m)}_{m_{1}m_{2}}(x) =
\frac{1}{2^{2j-1}} \sum_{\ell_{1}}^{}\sum_{\ell_{2}}^{}
\Gamma (j,m_{1},m,\ell_{1})\Gamma(j,m_{2},m,\ell_{2})
\nonumber\\
&& \quad \times\sum_{k_{1}=0}^{A^{(j, m, m_1)}_{\ell_{1},\ell_{2}}} 
\sum_{k_{2}=0}^{B^{(m,m_2)}_{\ell_{1},\ell_{2}}} 
{A^{(j, m, m_1)}_{\ell_{1},\ell_{2}}\choose k_{1}} 
{B^{(m,m_2)}_{\ell_{1},\ell_{2}}\choose k_{1}}
(-1)^{k_{1}} x^{k_{1}+k_{2}} f_{\tau}^{(m_{2}-m_{1})}(x) 
e^{-i(m_{2}-m_{1})\gamma}.
\label{eqn:matrix}
\end{eqnarray}
Here the summations $\sum_{\ell_{1}}^{}$ and 
$\sum_{\ell_{2}}^{}$ extend over all integers
of $\ell_{1}$ and $\ell_{2}$,
for which the arguments of the factorials
are positive or null, 
$$
A^{(j, m, m_1)}_{\ell_{1},\ell_{2}}=2j-(m-m_{1})-(\ell_{1}+\ell_{2}),
\quad 
B^{(m,m_2)}_{\ell_{1},\ell_{2}}=(m-m_{2})+(\ell_{1}+\ell_{2}),
$$
and
\begin{eqnarray}
f_{\tau}^{(a)}(x)=\sum_{k_{0}=0}^{[a/2]}
\sum_{k_{1}=0}^{k_{0}}\sum_{k_{2}=0}^{k_{1}}
{a \choose 2k_{0}}  {k_{0} \choose k_{1}}  {k_{1} \choose k_{2}}          
(-1)^{k_{0}+k_{1}} 
\tau^{a-2(k_{0}-k_{2})} x^{a-2(k_{0}-k_{1})}
\label{eqn:f_tau}
\end{eqnarray}
with $\tau = \tan(\beta/2)$, 
where $[z]$ denotes the integer not greater than $z$.
For example,
$f_{\tau}^{(1)}(x)=\tau x, f_{\tau}^{(2)}(x)=(2\tau^2+1)x^{2}-1,
f_{\tau}^{(3)}(x)=(4\tau^{3}+3\tau)x^{3}-3\tau x,
f_{\tau}^{(4)}(x)=(8\tau^{4}+8\tau^{2}+1)x^{4}-(8\tau^{2}+2)x^{2}+1$.
Note that $f_{\tau}^{(a)}(x)$ is even (respectively, odd)
if $a$ is even (respectively, odd).
The derivation of (\ref{eqn:matrix}) is tedious but
straightforward following the method given in \cite{KFK05,MKK07}.
The key formulas are found in Appendix C of \cite{MKK07}.
Combination of the expression (\ref{eqn:matrix}) with 
the symmetry properties (\ref{eqn:hermitian}) and (\ref{eqn:symmetry}) 
determines all elements of the matrix ${\bf M}^{(j,m)}(x)$ 
for any given $j \in \{1/2,1,3/2,\cdots\}$ and
$m \in \{-j, -j+1, \cdots, j\}$. 
%%%%%%%%%%%%%%%%%%%%%%%%%%%%%%%%%%%%%%%%%%%%%%%%%%%%%%%%%%%%%%%%%%%%%

\subsection{Recurrence formulas}
%%%%%%%%%%%%%%%%%%%%%%%%%%%%%%%%%%%%%%%%%%%%%%%%%%%%%%%%%%%%%%%%%%%%%%
As shown in Appendix B, from our expression (\ref{eqn:matrix}), 
we can derive the 
following recurrence formula for matrix elements 
${\bf M}^{(j,j)}_{m_{1}m_{2}}(x)$,
when the condition (\ref{eqn:condition1}) is satisfied,
\begin{eqnarray}
{\bf M}^{(j,j)}_{m_{1}m_{2}}(x) &=& \frac{1}{2}c(j;m_{1},m_{2})
(1-x){\bf M}^{(j-1/2, j-1/2)}_{m_{1}-1/2 \, m_{2}-1/2}(x)
{\bf 1}_{\{m_{1}\neq -j\}} \nonumber\\
&& \quad +\frac{1}{2^{2j-1}} f^{(2j)}_{\tau}(x)
e^{-2ij \gamma} {\bf 1}_{\{m_{1}=-j, m_2=j \}},     
\label{eqn:rec_1}                 
\end{eqnarray}
where
$$
c(j;m_{1},m_{2})=\frac{2j}{ \sqrt{(j+m_1)(j+m_2)} }.
$$
By solving this recurrence formula under the initial condition
(\ref{eqn:A1}),
and by using the symmetry properties (\ref{eqn:hermitian}) 
and (\ref{eqn:symmetry}),
matrices ${\bf M}^{(j,j)}(x)$ can be easily calculated 
even for large $j$.
Moreover, (\ref{eqn:matrix}) gives the following relation,
\begin{equation}
{\bf M}^{(j,j-1)}_{m_{1}m_{2}}(x)=
\frac{2(jx+m_{1})(jx+m_{2})}{j(1-x)(1+x)}{\bf M}^{(j,j)}_{m_{1}m_{2}}(x),
\label{eqn:rec_2}
\end{equation}
which enables us to determine
${\bf M}^{(j,j-1)}(x)$ from ${\bf M}^{(j,j)}(x)$.

%%%%%%%%%%%%%%%%%%%%%%%%%%%%%%%%%%%%%%%%%%%%%%%%%%%%%%%%%%%%%%%
\section{Analysis of Limit Distributions for Large $j$}
%%%%%%%%%%%%%%%%%%%%%%%%%%%%%%%%%%%%%%%%%%%%%%%%%%%%%%%%%%%%%%%
\subsection{Numerical evaluation of exact formula}
%%%%%%%%%%%%%%%%%%%%%%%%%%%%%%%%%%%%%%%%%%%%%%%%%%%%%%%%%%%%%%%
Though the formula (\ref{eqn:matrix}) with (\ref{eqn:f_tau}) 
is rather complicated, it is exact 
for any given values of $j,m,m_{1}$ and $m_{2}$.
Therefore, if we fix the value of $x$, it is easy to evaluate 
${\bf M}^{(j,m)}_{m_{1}m_{2}}(x)$ with any precision using computer.
In order to demonstrate validity of this procedure,
here we compare the numerical evaluations of our exact formula 
(\ref{eqn:matrix})
with the results of direct computer-simulations \cite{MKK07}
of the quantum-walk models
for a relatively large value of $j$.
As an example, we set
$j=11/2$ and $(\alpha,\beta,\gamma)=(0,\pi/2,\pi)$.
The initial qudit has $2j+1=12$ complex components.
Figure 1(a) shows the result, when we choose the initial qudit as
$\phi_{0}={}^{T}((1+i)/2,0,0,0,0,0,0,0,0,0,0,(1-i)/2)$,
where the thick lines show the exact limit-distribution of 
pseudovelocity $\nu^{(11/2)}(v)$ obtained by the above mentioned 
numerical calculation and 
the scattering dots indicate the distribution of $X_{t}/t$
at time step $t=100$ obtained 
by direct computer-simulation.
For this initial qudit, the limit probability-density 
$\nu^{(11/2)}(v)$ is symmetric and well describes the distribution
of $X_{t}/t$ for large $t$.
If we choose the initial qudit as 
$\phi={}^{T}(1+i,0,1+i,1,i,i,1+i,i,i,1+i,i,1+i)$
the distribution becomes asymmetric as shown by Fig.1(b).
%%%%%%%%%%%%%%%%%%%%%%%%%%%%%%%%%%%%%%%%%%%%%%%%%%%%%%%%
\begin{figure}[htpb]
\includegraphics[width=1.0\linewidth]{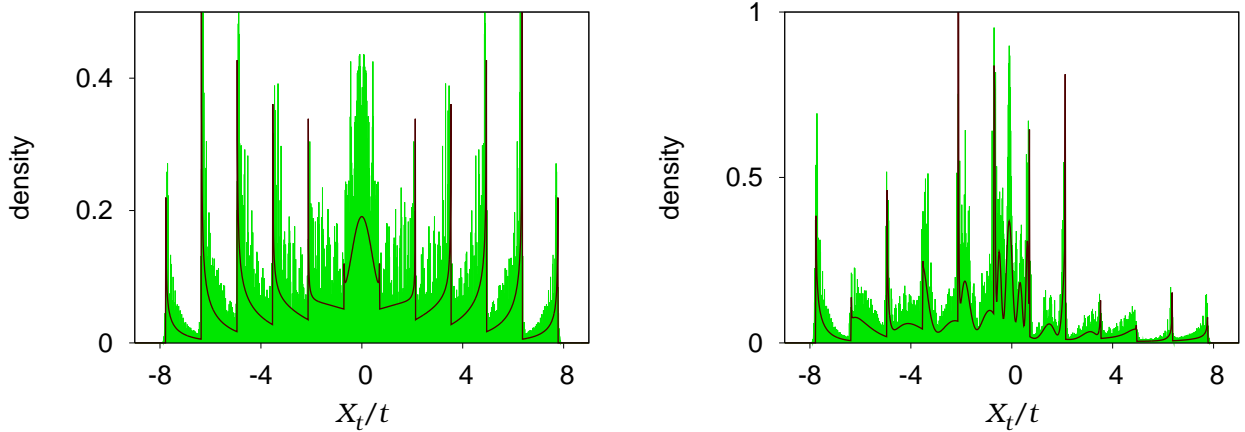}
\caption{(Color online)
Comparison between direct simulation results
and the probability densities of 
limit distributions for the twelve-component
model. (a) Symmetric and (b) asymmetric cases.
\label{eqn:3A}}
\end{figure}
%%%%%%%%%%%%%%%%%%%%%%%%%%%%%%%%%%%%%%%%%%%%%%%%%%%%%%%%

Figure 1(a) shows that the limit distribution for the 
$j=11/2$ model (twelve-component model) is given by
superposition of six Konno's density functions (\ref{eqn:Konno}),
each of which is appropriately scaled 
according to Eq.(\ref{eqn:dist}).
In addition to them, 
inside of the innermost Konno's density function,
we can see a convex structure in Fig.1(a)

%%%%%%%%%%%%%%%%%%%%%%%%%%%%%%%%%%%%%%%%%%%%%%%%%%%%%%%%%%%%%%%
\subsection{The convex structure}
%%%%%%%%%%%%%%%%%%%%%%%%%%%%%%%%%%%%%%%%%%%%%%%%%%%%%%%%%%%%%%%
From now on we fix the parameters of 
quantum coins as $\alpha=\gamma=0$
and the form of the initial qudits as 
$\phi_{0}={}^{T}(q,0,\cdots,0,\bar{q})$ with $q=(1+i)/2$.
In order to avoid Dirac's delta-function peaks, we will assume
that the number of states $2j+1$ is even.
Now we see the $j$-dependence of 
central convex structures.
Figure 2 shows the central parts of limit distributions with
a fixed window, $X_{t}/t\in [-2,2]$,
for variety of $j$'s, when we set $\beta=\pi/2$.
When $2j+1\geq 8$, we see convex structures
for $\beta=\pi/2$.
%%%%%%%%%%%%%%%%%%%%%%%%%%%%%%%%%%%%%%%%%%%%%%%%%%%%%%%
\begin{figure}[htpb]
\includegraphics[width=0.5\linewidth]{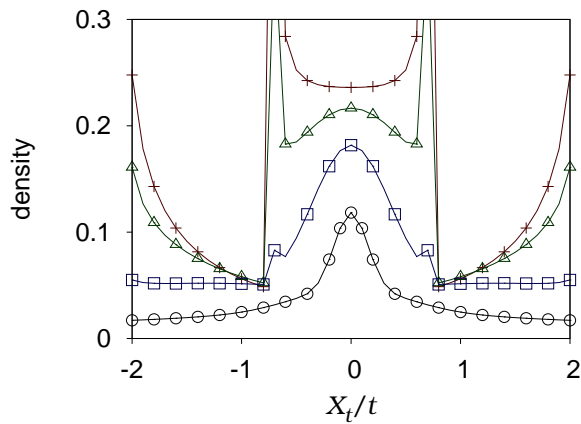}
\caption{(Color online)
Central parts of limit distributions for the models with
$2j+1=6$ (plotted by crosses),
8 (triangles), 14 (squares), and 50 (circles).
The convex structure appears around the origin,
when the number of states with $2j+1$ 
becomes greater than eight for $\beta=\pi/2$.}
\end{figure}
%%%%%%%%%%%%%%%%%%%%%%%%%%%%%%%%%%%%%%%%%%%%%%%%%%%%%%%

In order to verify the fact that, for any value of parameter $\beta$,
the convex-structure appears around the origin,
if the number of state $2j+1$ becomes sufficiently large,
we calculate the second derivative of the density of 
limit distribution (\ref{eqn:dist}).
By using the exact expression of weight functions given
in Subsection II.A, we have obtained the result,
\begin{eqnarray}
&& \frac{d^{2} \nu^{(j)}(v)}{dv^{2}}\Bigg|_{v=0}
= \frac{\sqrt{1-\cos^2 (\beta/2)}}{\pi \cos(\beta/2)}
\nonumber\\
&& \quad \times
\sum_{0<m\leq j}^{} \frac{1}{8 m^3}
\left[ \left\{2+\frac{1}{\cos^2 (\beta/2)}
+2(2m^{2}-j) \right\} 
\frac{(2j)!}{2^{2j-1} (j+m)!(j-m)!} \right].
\label{eqn:derivative_2}
\end{eqnarray}
%%%%%%%%%%%%%%%%%%%%%%%%%%%%%%%%%%%%%%%%%%%%%%%%%%%%%%%
\begin{figure}[htpb]
\includegraphics[width=0.5\linewidth]{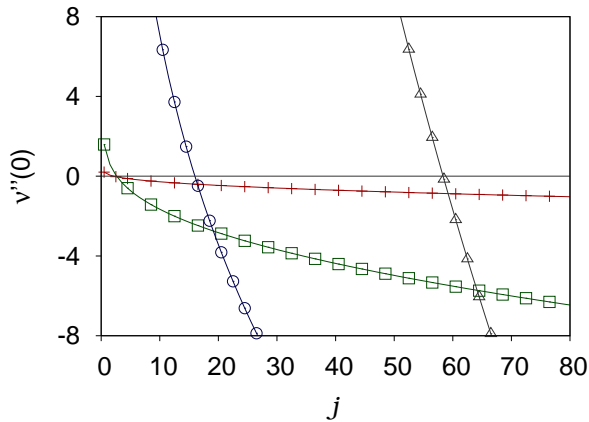}
\caption{(Color online)
Values of second derivatives of limit density-functions
at the origin are plotted for various $\beta$.
Crosses are for $\beta=\pi/10$, 
squares for $\beta=\pi/2$,
circles for $\beta=22 \pi/25$, 
and triangles for $\beta=46 \pi/50$,
respectively.}
\end{figure}
%%%%%%%%%%%%%%%%%%%%%%%%%%%%%%%%%%%%%%%%%%%%%%%%%%%%%%%

In Fig.3 we plot the values of (\ref{eqn:derivative_2})
with changing values of $j$ for $\beta=\pi/10,
\pi/2, 22\pi/25$ and $47\pi/50$.
As demonstrated by this figure, we can prove that 
for any given $\beta\in [0,\pi]$,
these is a critical value $j_{c}(\beta)$
such that $d^{2}\nu^{(j)}(v)/dv^{2}|_{v=0}< 0$ 
for all $j>j_{c}(\beta)$.

%%%%%%%%%%%%%%%%%%%%%%%%%%%%%%%%%%%%%%%%%%%%%%%%%%%%%%%%%%%%%%%
\subsection{Smoothing by weight function in large $j$}
%%%%%%%%%%%%%%%%%%%%%%%%%%%%%%%%%%%%%%%%%%%%%%%%%%%%%%%%%%%%%%%
As given by Eq.(\ref{eqn:dist}),
the probability density of limit distribution of 
pseudovelocities is given by superposing 
Konno's density functions (\ref{eqn:Konno})
appropriately scaled.
Since each Konno's density function $\mu(v/2m;\cos(\beta/2))$
has a finite open support $v\in (-2m\cos(\beta/2),2m\cos(\beta/2))$
and diverges at the boundaries of it, 
we see pikes at $v=\pm2m\cos(\beta/2)$,
$0< m\leq j$, in the limit distribution as 
shown by Fig.1(a),
which was given for the case $2j+1=12$ and $\beta=\pi/2$.

For a given value of parameter $\beta$, however,
if we set $j\gg j_{c}(\beta)$, limit distributions
seem to be quite different from that shown in Fig.1(a).
Figure 4(a) shows the limit distribution for the 
$2j+1=50$ case with $\beta=\pi/2$.
Pikes vanish in a central region and the convex structure
at the origin becomes very evident.
%%%%%%%%%%%%%%%%%%%%%%%%%%%%%%%%%%%%%%%%%%%%%%%%%%%%%%%
\begin{figure}[htpb]
\includegraphics[width=1.0\linewidth]{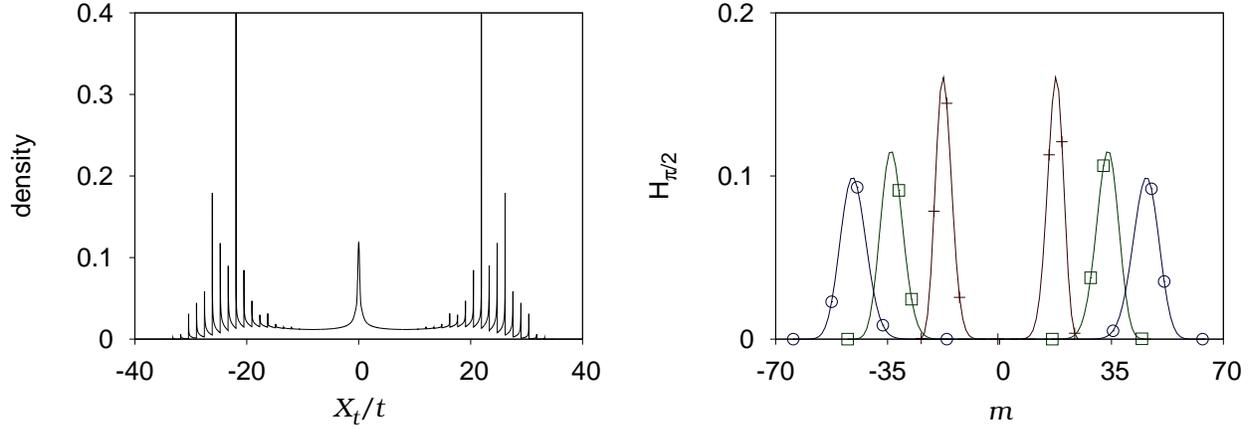}
\caption{(Color online)
(a) Probability density of 
limit distribution for the case $2j+1=50$, $\beta=\pi/2$.
(b) Functions $H_{\pi/2}(m)$ plotted for the cases
$2j+1=50$ (crosses), 96 (squares), and 130 (circles).
\label{fig:3B3}}
\end{figure}\\
%%%%%%%%%%%%%%%%%%%%%%%%%%%%%%%%%%%%%%%%%%%%%%%%%%%%%%%

This smoothing phenomenon is caused by the interesting 
property of weight functions
${\cal M}^{(j,m)}$; for large $j$,
${\cal M}^{(j,m)}(v/2m)$ becomes to attain zeros
at the points $v= \pm 2m \cos(\beta/2),
0 < m \leq j$,
where the scaled Konno's density functions,
$\mu(v/2m; \cos(\beta/2))$, diverge.
In order to see this fact,
for given $j$ and $\beta$, we define
a function of $m$ by
\begin{eqnarray}
H^{(j)}_{\beta}(m) &=& {\cal M}^{(j,m)}
(\cos \beta/2 )\nonumber\\
&=& \frac{(2j)!}{2^{2j-1}}
\sum_{k_{1}=0}^{j+m}\sum_{k_{2}=0}^{j-m}
\frac{(-1)^{k_{1}}
(\cos(\beta/2))^{k_{1}+k_{2}}}
{k_{1}!(j+m-k_{1})!k_{2}!(j-m-k_{2})!}
{\bf 1}_{\{k_1+k_2 \, \mbox{is even}\}},          
\nonumber
\end{eqnarray}
$0 < m \leq j$.
Figure 4(b) shows the functions, when $\beta=\pi/2$, 
for the cases $2j+1=50,96$ and $130$.
As increasing $j$, the central region, where $H_{\beta}^{(j)}(m)=0$, 
becomes wider.
%%%%%%%%%%%%%%%%%%%%%%%%%%%%%%%%%%%%%%%%%%%%%%%%%%%%%%%%%%%%%%%
\subsection{Rescaling of limit density-function}
%%%%%%%%%%%%%%%%%%%%%%%%%%%%%%%%%%%%%%%%%%%%%%%%%%%%%%%%%%%%%%%
By definition of our models, the range of elementary
hopping of quantum walker at each time step is $2j$;
see Eq.(\ref{eqn:hopp}).
Then distribution of pseudovelocities of quantum walker spreads in
an interval $(-2j\cos(\beta/2),2j\cos(\beta/2))$.

In order to discuss the $j \to \infty$ limit of the series of our models,
here we introduce the rescaled variable
\begin{equation}
\widetilde{X}_{t}^{(j)}=\frac{X_{t}^{(j)}}{2j\cos(\beta/2)}
\label{eqn:scale}
\end{equation}
for each value of $\beta$.
Figure 5(a) shows the limit density-functions of
the rescaled pseudovelocities $\widetilde{X}^{(j)}_{t}/t$ for
$2j+1=10, 20$ and $50$.
In this variable, the support of the limit 
distribution is fixed to be $(-1,1)$.
As shown by Fig.5(a), the central convex structure becomes 
sharper monotonically in increasing the value of $j$.
Corresponding to (\ref{eqn:scale}),
we plot $\sigma_{j}H_{\beta/2}$'s as functions of
$m/{\sigma_{j}}$,
where $\sigma_{j}=\sqrt{2}j$, in Fig.5(b) for $\beta=\pi/2$.
It is interesting to see that the locations,
where the functions take non-zero values, are now fixed.
%%%%%%%%%%%%%%%%%%%%%%%%%%%%%%%%%%%%%%%%%%%%%%%%%%%%%%%%%%%%%%
\begin{figure}[htpb]
\includegraphics[width=1.0\linewidth]{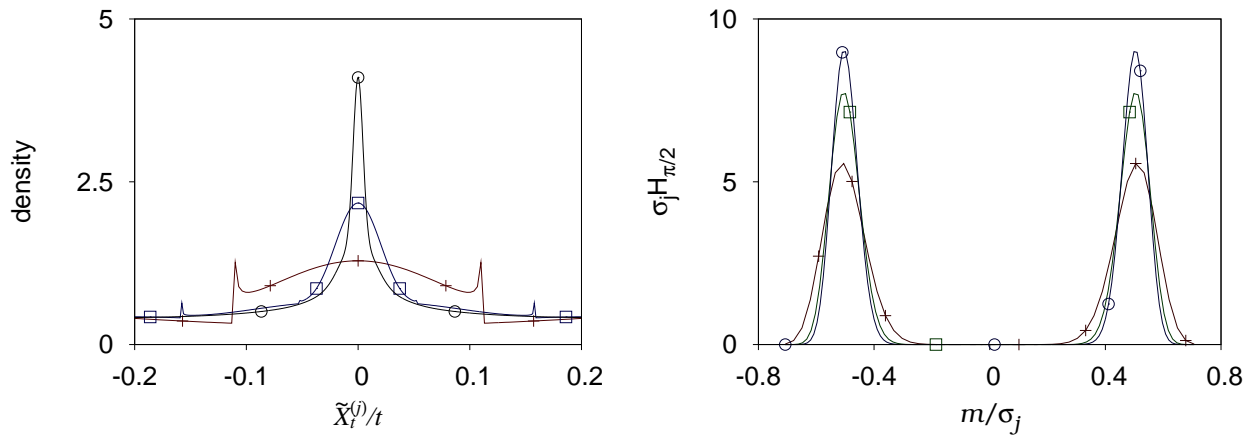}
\caption{(Color online)
(a) Central parts of density functions of limit distributions
of rescaled pseudovelocities for 
$2j+1=10$ (crosses), 20 (squares), and 50 (circles).
(b) $\sigma_j H_{\beta}(m)$ versus $m/\sigma_j$
with $\sigma_j=\sqrt{2}j$
for $2j+1=50$ (crosses), 96 (squares) and 130 (circles),
when $\beta=\pi/2$.}
\end{figure}
%%%%%%%%%%%%%%%%%%%%%%%%%%%%%%%%%%%%%%%%%%%%%%%%%%%%%%%%%%%%%%%

%%%%%%%%%%%%%%%%%%%%%%%%%%%%%%%%%%%%%%%%%%%%%%%%%%%%%%%%%%%%%%%
\section{Discussion}
%%%%%%%%%%%%%%%%%%%%%%%%%%%%%%%%%%%%%%%%%%%%%%%%%%%%%%%%%%%%%%%

Now we discuss the $j \to \infty$ limit of the
present series of one-dimensional quantum-walk models.
If we consider the situation that $j$ is finite but
very large, the limit distribution of the rescaled
pseudovelocity $\widetilde{X}^{(j)}_t/t$ will have
a simple profile.
There will be a sharp convex structure at the origin
and accumulations of pikes only in the very vicinity
of the boundary points -1 and 1 of the support
of limit density-function,
but in other regions in the support
the density of distribution will be almost zero.
In the $j \to \infty$ limit, the central
convex-structure will become a single point mass
({\it i.e.} a Dirac delta-function) at the origin.
It implies that the pseudovelocity is almost
zero, that is, the quantum walk will lose
its velocity in the $j \to \infty$ limit.

In the series of one-dimensional quantum-walk models
studied in the present paper,
we have used Wigner's rotation matrices $R^{(j)}$
as quantum coins.
We should note that $R^{(j)}$ is introduced \cite{Wig59,Mes62},
when rotations in the three-dimensional real space 
are quantized and the index $j$ is 
defined as a quantum number,
which specify physical states allowed in the
quantum mechanics.

In the standard quantum mechanics, a small but finite
parameter $\hbar$ (the Planck constant divided by $2\pi$)
is introduced and physical quantities are quantized
to have only discrete values of the form
$\hbar \times \mbox{(quantum numbers)}$.
For example the energy levels of a harmonic
oscillator with the angular frequency $\omega$
are given by $E_n = \hbar \omega (n+1/2),
n=0,1,2, \dots$,
and the square of angular momentum ${\bf L}$ and 
its $z$-component are given by
${\bf L}^2=\hbar^2 \ell(\ell+1)$ and
$L_z=\hbar m$,
$\ell=0,1/2,1,3/2, \cdots, m=-\ell, -\ell+1, \cdots, \ell$.
If the quantum system involves $\hbar$ explicitly,
it is easy to consider its classical correspondence
by taking the limit $\hbar \to 0$.
Even though the system does not include the
parameter $\hbar$ explicitly,
the classical limit can be realized
by taking a large quantum-number limit, since
physical quantities should be given of the form
$\hbar \times \mbox{(quantum numbers)}$
as the above examples show.

Then we can expect that, if we take $j \to \infty$
limit appropriately, the classical diffusive behavior
will be observed \cite{BCA03b}.
One possibility is to see a crossover phenomenon
from quantum-walk behavior to classical diffusion
in $ j \to \infty$.
Assume the form $X_t^{(j)} \sim j t F(t/j^{\theta})$
with a scaling function $F(z)$ such that
$F(z) \sim z^{-1/2}$ in $z \to 0$ and
$F(z) \to$ const. in $z \to \infty$, where 
$\theta$ is an exponent.
For finite $j$, $X_t^{(j)} \sim j t$
in $t \to \infty$, as we have shown in the
present paper.
On the other hand, for finite $t$, we will have
$X_t^{(j)} \sim (jt)(t/j^{\theta})^{-1/2}
=j^{1+\theta/2} \sqrt{t}$ in $j \to \infty$;
that is, 
$X^{(j)}_{t}/j^{1+\theta/2}$ is diffusive
in large $t$.
It will be an interesting future problem to
clarify the phenomena, which are realized
when we take a proper classical limit
in the series of quantum-walk models.

%%%%%%%%%%%%%%%%%%%%%%%%%%%%%%%%%%%%%%%%%%%%%%%%%%%%%%%%%%%%
\begin{acknowledgments}
This works is partially supported by the Grand-in-Aid for
scientific research (KIBAN-C, No. 17540363) of 
Japan society for the promotion of science.
\end{acknowledgments}

%%%%%%%%%%% APPENDICES %%%%%%%%%%%%%%%%%%%%%%%%%%%%%%%%
\appendix
%%%%%%%%%%%%%% Appendix A %%%%%%%%%%%%%%%%%%%%%%%%%%%%%%%%%%%%%%%
\section{The matrices ${\bf M}^{(j,m)}(x)$ for $j=1/2, 1, 3/2$}%%
%%%%%%%%%%%%%%%%%%%%%%%%%%%%%%%%%%%%%%%%%%%%%%%%%%%%%%%%%%%%%%%%%
Let $\tau=\tan(\beta/2)$ and 
$f_{\tau}^{(1)}(x)=\tau x, f_{\tau}^{(2)}(x)=(2\tau^2+1)x^{2}-1$.
Then
\begin{eqnarray}
\label{eqn:A1}
&& {\bf M}^{(1/2,1/2)}(x) = \left(\begin{array}{ll}
            1-x & \tau x e^{i\gamma} \\
            \tau xe^{-i\gamma} & 1+x \\            
          \end{array}\right), \\
&& 
{\bf M}^{(1,1)}(x) = \left(\begin{array}{lll}
\frac{1}{2}(1-x)^2  & \frac{\sqrt{2}}{2}(1-x)f_{\tau}^{(1)}(x) e^{i\gamma} 
& \frac{1}{2}f_{\tau}^{(2)}(x) e^{2i\gamma}\\
\frac{\sqrt{2}}{2}(1-x)f_{\tau}^{(1)}(x) e^{-i\gamma} 
& (1-x)(1+x) & \frac{\sqrt{2}}{2}(1+x)f_{\tau}^{(1)}(x) e^{\gamma}\\
\frac{1}{2}f_{\tau}^{(2)}(x) e^{-2i\gamma} 
& \frac{\sqrt{2}}{2}(1+x)f_{\tau}^{(1)}(x) e^{-i\gamma} 
& \frac{1}{2}(1+x)^2\\
\end{array}\right), \nonumber
\end{eqnarray}
\begin{footnotesize}
\begin{eqnarray}
&&{\bf M}^{(3/2,3/2)}(x) \nonumber\\
&=&\left(\begin{array}{llll}
\frac{1}{4}(1-x)^3  & \frac{\sqrt{3}}{4}(1-x)^2f_{\tau}^{(1)}(x) e^{i\gamma} 
& \frac{\sqrt{3}}{4}(1-x)f_{\tau}^{(2)}(x) e^{2i\gamma} 
& \frac{1}{4}f_{\tau}^{(3)}(x) e^{3i\gamma} \\
\frac{\sqrt{3}}{4}(1-x)^2f_{\tau}^{(1)}(x) e^{-i\gamma} 
& \frac{3}{4}(1-x)^2(1+x) 
& \frac{3}{4}(1-x)(1+x)f_{\tau}^{(1)}(x) e^{i\gamma} 
& \frac{\sqrt{3}}{4}(1+x)f_{\tau}^{(2)}(x) e^{2i\gamma} \\
\frac{\sqrt{3}}{4}(1-x)f_{\tau}^{(2)}(x) e^{-2i\gamma} 
& \frac{3}{4}(1-x)(1+x)f_{\tau}^{(1)}(x) e^{-i\gamma} 
& \frac{3}{4}(1-x)(1+x)^2 
& \frac{\sqrt{3}}{4}(1+x)^2f_{\tau}^{(1)}(x) e^{i\gamma} \\
\frac{1}{4}f_{\tau}^{(3)}(x) e^{-3i\gamma} 
& \frac{\sqrt{3}}{4}(1+x)f_{\tau}^{(2)}(x) e^{-2i\gamma} 
& \frac{\sqrt{3}}{4}(1+x)^2f_{\tau}^{(1)}(x) e^{-i\gamma} 
& \frac{1}{4}(1+x)^3 \\
\end{array}\right). \nonumber
\end{eqnarray}
\end{footnotesize}
%%%%%%%%%%%%%% Appendix B %%%%%%%%%%%%%%%%%%%%%%%%%%%%%
\section{Derivation of EQ.(\ref{eqn:rec_1})}%%%%%%%%%%%%%%%
%%%%%%%%%%%%%%%%%%%%%%%%%%%%%%%%%%%%%%%%%%%%%%%%%%%%%%%
When $m_{1}\not= -j$,
Eq.(\ref{eqn:matrix}) gives
\begin{eqnarray}
&& (1-x){\bf M}^{(j-1/2, j-1/2)}_{m_{1}-1/2 \, m_{2}-1/2}(x) 
=\frac{1}{2^{2j-2}}
f_{\tau}^{(m_{2}-m_{1})}(x) e^{-i(m_{2}-m_{1})\gamma} \nonumber\\
&& \quad \times 
\Gamma\left(j-\frac{1}{2},m_{1}-\frac{1}{2},j-\frac{1}{2},0 \right) 
\Gamma \left(j-\frac{1}{2}, m_{2}-\frac{1}{2},
j-\frac{1}{2}, 0 \right) \nonumber \\
&& \quad \times \sum_{k_{1}=0}^{j+m_{1}-1}
\sum_{k_{2}=0}^{j-m_{2}} 
{j+m_{1}-1 \choose k_{1}}        
{j-m_{2} \choose k_{2}} 
(-1)^{k_{1}} (x^{k_{1}+k_{2}}-x^{k_{1}+k_{2}+1}) \nonumber.
\end{eqnarray}
The summations over $k_{1}$ and $k_{2}$ are carried out as 
\begin{eqnarray}
&& \sum_{k_{2}=0}^{j-m_{2}} 
{j-m_{2} \choose k_{2}}x^{k_{2}}
\left\{\sum_{k_{1}=0}^{j+m_{1}-1} 
{j+m_{1}-1 \choose k_{1}}(-1)^{k_{1}}x^{k_{1}} \nonumber
-\sum_{k_{1}=0}^{j+m_{1}-1} 
{j+m_{1}-1 \choose k_{1}}(-1)^{k_{1}}x^{k_{1}+1} \right\}
\nonumber\\
&& \quad =\sum_{k_{2}=0}^{j-m_{2}} 
{j-m_{2} \choose k_{2}}x^{k_{2}}
\left\{1+(-x)^{j+m_{1}}
+\sum_{k_{1}=1}^{j+m_{1}-1}
{j+m_{1} \choose k_{1}}(-1)^{k_{1}}x^{k_{1}} \right\} \nonumber\\
&& \quad = \sum_{k_{1}=0}^{j+m_{1}}\sum_{k_{2}=0}^{j-m_{2}} 
{j+m_{1} \choose k_{1}}
{j-m_{2} \choose k_{2}} (-1)^{k_{1}} x^{k_{1}+k_{2}}.
\nonumber
\end{eqnarray}
We note the equality
\begin{eqnarray}
&& c(j;m_{1},m_{2})
\Gamma\left(j-\frac{1}{2}, m_{1}-\frac{1}{2}, 
j-\frac{1}{2}, 0 \right) 
\Gamma\left(j-\frac{1}{2}, m_{2}-\frac{1}{2}, 
j-\frac{1}{2}, 0 \right) \nonumber\\
&& \quad = 
\Gamma(j,m_{1},j,0)\Gamma(j,m_{2},j,0).
\nonumber
\end{eqnarray}
Then Eq.(\ref{eqn:rec_1}) is derived.
When $m_{1}=-j$, it is enough to only consider the
case $m_{2}=j$
under the condition (\ref{eqn:condition1}).
This case is trivial;
${\bf M}^{(j,m)}_{-jj}(x)=
2^{-2j+1}f_{\tau}^{(2j)}(x)e^{-2i j\gamma}$.
%%%%%%%%%%%%%%%%%%%%%%%%%%%%%%%%%%%%%%%%%%%%%%%%%%%%%%%%%%%%

%%%%%%%%%%%%%%%%%%%%%%%%%%%%%%%%%%%%%%%%%%%%%%%%%%%%
%%%%%%%%%%%%%%%%%%%%%%%%%%%%%%%%%%%%%%%%%%%%%%%%%%%%
% Create the reference section using BibTeX:
%\bibliography{basename of .bib file}

\end{document}